# On the Convergence of the Current-Constrained Power Angle Curve of Virtual Admittance-Based Grid Forming Converters

Haoxiang Zong, *Member, IEEE*, Chen Zhang, *Member, IEEE*, Yu Zhang, *Member, IEEE*, Xu Cai, *IEEE,* and Marta Molinas, *Fellow, IEEE*

***Abstract*—The current-constrained power-angle curve (PAC) is crucial for the transient synchronization stability (TSS) analysis of virtual admittance-based (VA) grid-forming (GFM) converters. Its formulation and application rely on the quasi-steady-state assumption critically, i.e., the active power can converge to its steady-state across the entire angle space in both a stable and fast manner. Despite this assumption is intuitively perceivable, it lacks sufficient clarification, particularly on underlying behaviors if violated. To this end, this paper uncovers a new phenomenon on the non-uniform convergence of the VA-PAC. To achieve this, an eigen-sweep-based analysis of the full-order VA-PAC model with detailed controls is conducted, by which the existence of this issue is theoretically confirmed. On this basis, the open-loop stability and response-rate conditions of the full-order VA-PAC model for ensuring its convergence are clarified. Findings of this work can provide deeper insights into the existing TSS analyses of GFM converters, and are expected to provoke new analyses.**

## I. Introduction

GRID-forming (GFM) converters are regarded as one promising technique for operating the renewable power generations (RPGs)-dominated power systems. Given its significance, dynamic issues of the GFM converter, ranging from the grid support function to the transient synchronization stability (TSS), etc., have been widely studied. Among these, the TSS, as a fundamental issue in power system operation, has attracted particular attention lately.

Analysis of the GFM converter's TSS issue is closely related to that of synchronous generator (SG), as the GFM converter is intentionally designed to mimic the dominant behavior of the SG. Under this context, a series of SG-inspired TSS analysis methods, e.g., the equal area criterion (EAC) [1]-[3], the energy function [4], etc., have been developed. For these methods, the power angle curve (PAC) serves as an essential tool. However, the PAC of the GFM converter is fundamentally different from that of the SG due to the current limitation effect. Once being activated, an abrupt change in the PAC will occur, depending on the limitation strategy. For example, GFM converters' PACs using the circular-based and the *d*-axis priority-based limitation strategies are reported in [4] and [5], etc., respectively.

Although these PACs [4], [5] are well developed, they rely on quasi-steady-state assumptions. This means, these PACs are valid if and only if the dynamics associated with the circuit and inner-loop current control are adequately fast. However, there lacks a formal clarification on the conditions for which the actual PAC can converge to its steady-state. While, apparently, this clarification is significant for understanding the existing PAC-based TSS analyses' applicability and limitations. To fill this gap, this letter will reveal a new phenomenon associated with the convergence issue of the PAC along with a formal theoretical justification. To the authors' best knowledge, this is the first attempt to demonstrate this issue.

## II. The Non-Uniform Convergent Phenomenon

To demonstrate this issue without sacrificing generality, a GFM control with the virtual admittance-based (VA) votlage control, the circular-based current limitation, and the virtual SG-based synchronization control, is adopted, as in Fig. 1(a). The VA-PAC under this control scheme can be modeled as:

$$P(\delta) = \mathrm{Re}\left\{ \boldsymbol{U}_{\mathrm{s}}(\delta) \boldsymbol{I}_{\mathrm{s}}^{*}(\delta) \right\} \tag{1}$$

$$\boldsymbol{I}_{\mathrm{s}}(\delta) = \frac{\boldsymbol{E} - \boldsymbol{U}_{\mathrm{g}}}{k(\delta)\boldsymbol{Z}_{\mathrm{v}} + \boldsymbol{Z}_{\mathrm{g}}},\ \boldsymbol{U}_{\mathrm{s}}(\delta) = \boldsymbol{Z}_{\mathrm{g}}\boldsymbol{I}_{\mathrm{s}}(\delta) + \boldsymbol{U}_{\mathrm{g}} \tag{2}$$

where $\boldsymbol{Z}_{\mathrm{v}}$ is the virtual admittance; $\boldsymbol{Z}_{\mathrm{g}}$, $\boldsymbol{U}_{\mathrm{g}}$ are the impedance and voltage of the ac grid; $\boldsymbol{E}$ is the internal voltage; $P$ and $\delta$ are the active power and power angle. The angle-dependent scaling factor $k(\delta)$ is to account for circular-based limitation effects, i.e., via $k(\delta)\cdot\boldsymbol{Z}_{\mathrm{v}}$, where $k(\delta)>1$ if limitation is activated. Please refer to [4] for details on the VA-PAC modeling.

Derivation of (1) assumes an ideal voltage control, meaning that the circuit can settle into its steady-state stably and rapidly. This assumption is achievable, but under certain conditions. Fig. 1(b) shows the impact of the current loop bandwidth $f_{\mathrm{c}}$ on the convergence of the VA-PAC. When $f_{\mathrm{c}}$ decreases, the measured VA-PAC (denoted by PAC-sim) starts oscillating and deviating from the ideal one characterized by (1) (denoted by PAC-ana). Interestingly, this oscillatory deviation occurs mostly when the

This work was supported by the National Natural Science Foundation of China under Grant 52577207. *(Corresponding author: Chen Zhang).*

Haoxiang Zong, Chen Zhang, Yu Zhang and Xu Cai are with the School of Electrical Engineering, Shanghai Jiao Tong University, Shanghai, China. (e-mail: nealbc@sjtu.edu.cn). Marta Molinas is with the Department of Engineering Cybernetics, Norwegian University of Science and Technology, Trondheim, Norway. (e-mail: marta.molinas@ntnu.no).

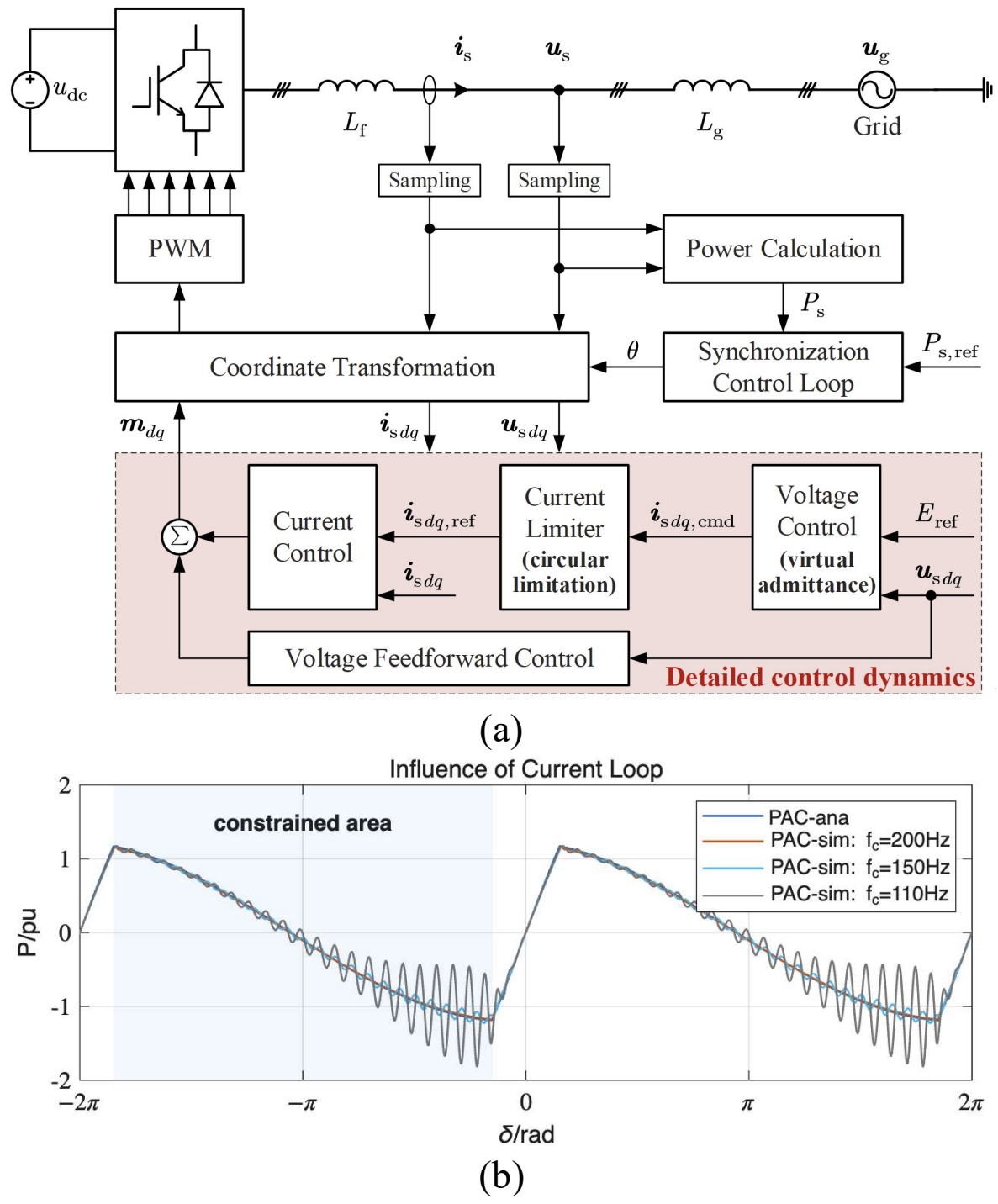

**Fig. 1.** Illustration of the non-uniform convergent phenomenon, (a) control scheme of the VA-GFM. (b) non-uniform convergence of the VA-PAC. (PAC-sim is measured from the simulation model, by replacing the synchronization loop with a constant angle-driven speed $\Delta\omega$. Here, $\Delta\omega$ is kept small, its impact will be discussed later.)

PAC enters into the current constrained area. This phenomenon is thereby referred to as the *non-uniform convergent* issue of the VA-PAC. Other controls, e.g., virtual admittance, will exhibit similar effects and are omitted here due to paper's length.

To uncover the presence of this phenomenon and evaluate its condition, a full order VA-PAC model incorporating inner-, outer-loop control and circuit dynamics is required, which will be presented and analyzed next.

## III. Full-Order VA-PAC Modeling And Convergence Condition Analysis

### A. *Establishment of the Full-Order VA-PAC Model*

The basic idea of its derivation is as follows: first, establish the detailed model of overall system; then, by selecting $\Delta P$ and $\Delta\omega$ as the input and output, the full-order VA-PAC can be obtained as $H_{sync}(s)$; afterwards, by piecing it together with the synchronization control path, a closed-loop model from the view-angle of synchronization can be obtained as Fig. 2(a).

Detailed derivation of $H_{sync}(s)$ relies on the linearization, which is a mature and well-documented tool [6], thus being omitted here. One thing worth noting herein is that, due to the activation of the current limiter, $H_{sync}(s)$ will undergo a model switch. When the limiter is not activated, the reference and actual command in Fig.1(a) are the same, see the first line of (3). Otherwise, the second line of (3) should be used to account for the small signal dynamics of the circular-based limiter. It is noted that for the considered VA-based voltage control, the current reference is continuous across the current-limiting switching point, thus enabling the piecewise modeling in (3).

$$\Delta \boldsymbol{i}_{sdq,\text{ref}} = \begin{cases} \Delta \boldsymbol{i}_{sdq,\text{cmd}} & k(\delta)=1 \\ \dfrac{\Delta \boldsymbol{i}_{sdq,\text{cmd}}}{k(\delta) I_{\lim 0}^2} \begin{bmatrix} I_{\lim 0}^2 - I_{d0}^2 & -I_{d0}I_{q0} \\ -I_{d0}I_{q0} & I_{\lim 0}^2 - I_{q0}^2 \end{bmatrix} & k(\delta)>1 \end{cases} \tag{3}$$

where $I_{\lim0}$ is the preset current limit value, and $I_{d0}$, $I_{q0}$ are the steady-state values of the $dq$ currents behind the current limiter, i.e., the actual command, and at a specific angle $\delta$.

The parameterized model structure of $H_{sync}(s)$ is shown in (4). It can be seen that detailed control and circuit dynamics in Fig. 1(a) are incorporated, including virtual admittance loop $K_{vir}(s)$, current loop $K_{cc}(s)$, voltage feedforward loop $K_f(s)$, ac gird inductance $L_g$, ac filter $L_f$, and steady-state points $\boldsymbol{\rho}_0(\delta)$. Also, The limitation scaling factor $k(\delta)$ is used to account for the model switching in (3).

$$H_{\text{sync}}\left(K_{\text{vir}}(s), K_{\text{cc}}(s), K_{\text{f}}(s), L_{\text{g}}, L_{\text{f}}, k(\delta), \boldsymbol{\rho}_0(\delta)\right) \tag{4}$$

Once $H_{sync}(s)$ is established, the convergence of the full-order VA-PAC can be evaluated by its poles. For example, if some of the poles fall into the right-hand plane (RHP), the PAC will present oscillation divergence as shown in Fig. 2(b). Two typical cases (i.e., Case I/II for unconstrained/constrained regions) are tested for validation. The poles distribution of $H_{sync}(s)$ under two cases is given in Table I, where it indicates the convergence in Case I and oscillatory divergence in Case II due to the absence/presence of RHP poles, respectively. These theoretical results agree well with the time-domain simulations in Fig. 2(b), and thus validating the correctness of (4).

### B. *Convergence Condition Analysis of the VA-PAC*

With the full-order VA-PAC (i.e., $H_{sync}(s)$), the conditions for meeting quasi-steady-state assumptions in VA-PAC can now be more formally and specifically interpreted as: 1) the stability of $H_{sync}(s)$ should be assured over the entire angle space; and 2) the response rate of $H_{sync}(s)$ should be sufficiently fast compared to the timescale of the synchronization control (i.e., the forward path in Fig. 2(a)) that governs the speed of the angle movement. These two influential conditions are analyzed more in detail as follows.

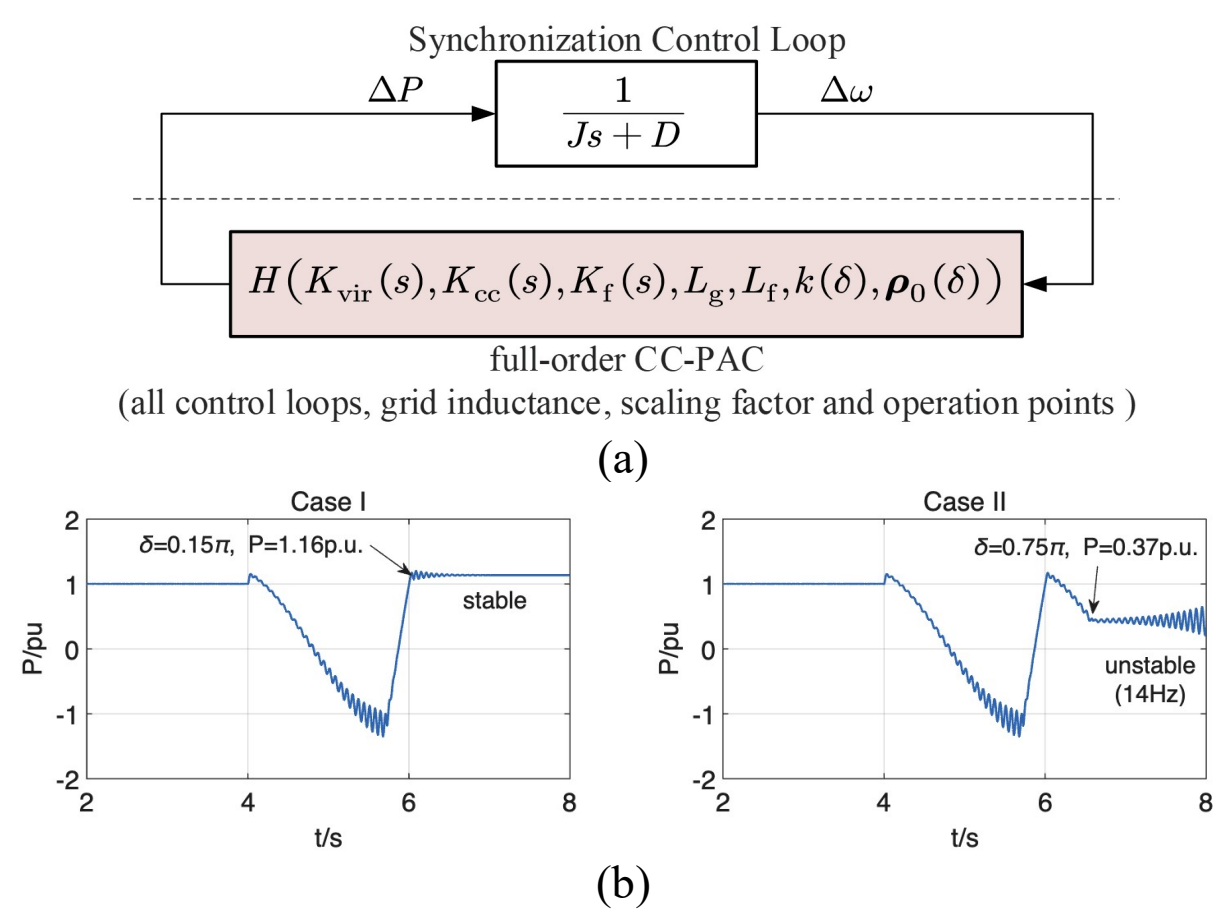

**Fig. 2.** The model structure and validation of the full-order VA-PAC. (a) synchronization control as the forward and full-order VA-PAC as the feedback. (b) time-domain verification.

TABLE I Poles Distribution of the Full-Order VA-PAC

| Case I | Poles | Case II | Poles |
|---|---|---|---|
| unconstrained point ($\delta\approx 0.15\pi$、$P\approx$ 1.16 p.u.) | **−19.36 ± j129.54**<br>−44.97 ± j20.76<br>−102 ± j241.75<br>−250.08<br>−249.99 | constrained point ($\delta\approx 0.75\pi$、$P\approx$ 0.37 p.u.) | **1.044 ± j85.40**<br>−55.94 ± j23.31<br>−86.30 ± j251.24<br>−242 + j39.79<br>−242 – j39.79 |

*1) Impact of Stability Condition on the Convergence*

Based on the full-order VA-PAC model $H_{sync}(s)$, Fig. 3(a) presents an eigen-sweep analysis by tracking the dominant eigenvalue over the entire angle space, visually reconstructing the first-row data in Table I. Specifically, in the unconstrained area, all eigenvalues meet $\mathrm{Re}\{\lambda\}<0$ condition, ensuring the stability of $H_{sync}(s)$ and the convergence of the PAC. In contrast, as the power angle moves into the constrained area, $\mathrm{Re}\{\lambda\}>0$ appears, indicating the violation of the stability condition and the PAC cannot converge to its steady-state in this area. This process well explains the event in Fig. 1 (under $f_c$ = 110Hz), where the PAC is initially converged to its steady-state, then diverges after entering the constrained area.

The above analysis demonstrates how the stability condition affects the uniform convergence of the VA-PAC. However, the quasi-steady-state assumption of the VA-PAC requires not only stability but also a sufficient response rate. This is because the VA-PAC in (1) depends on the angle dynamics, requiring fast convergence within the synchronization-control timescale. The impact of this condition is discussed next.

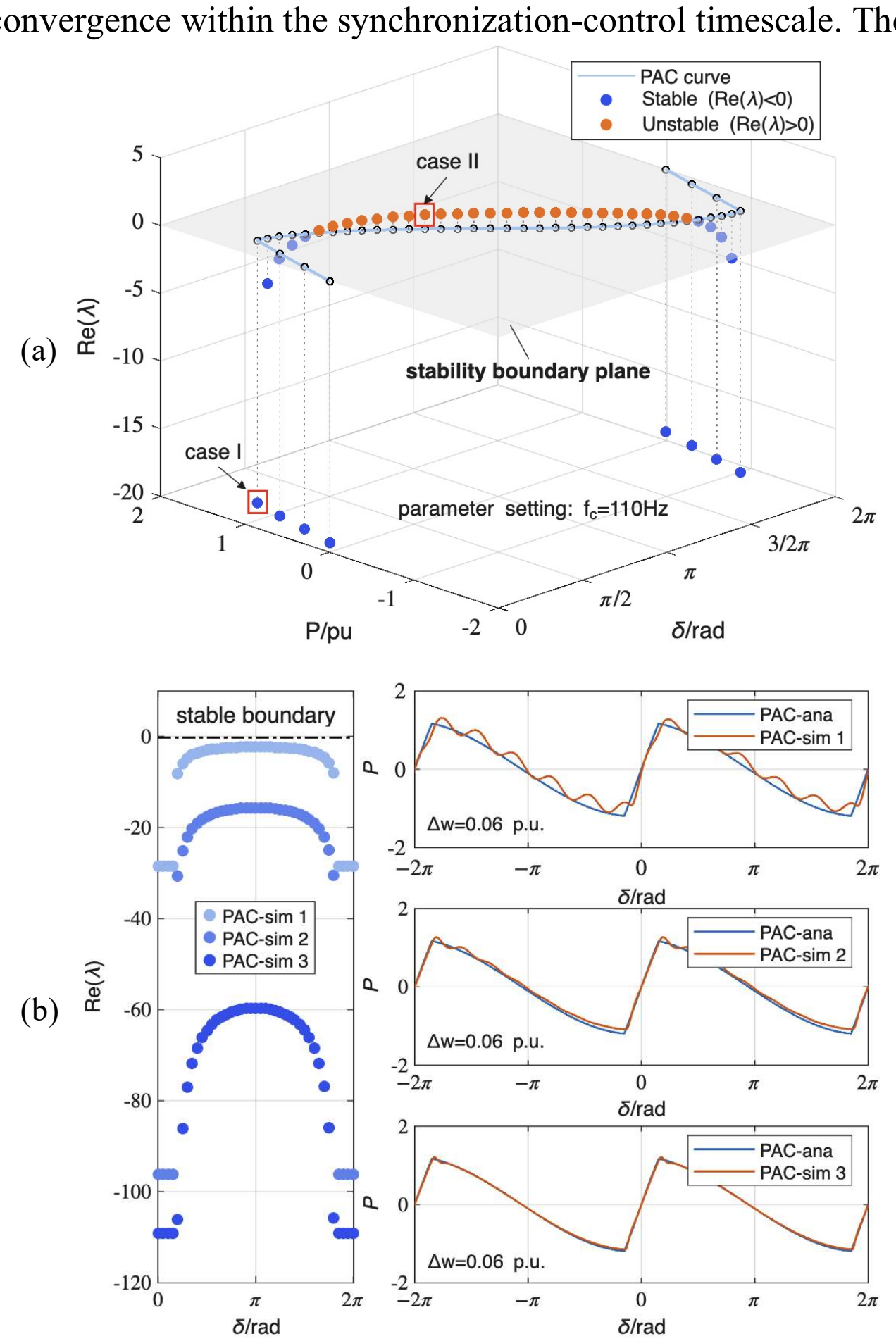

**Fig. 3.** Convergence condition analysis, (a) the eigen-sweep analysis of $H_{sync}(s)$ over the entire angle space, where $\mathrm{Re}\{\lambda\}$ is plotted along the $z$-axis, and the $x$-$y$ plane represents the $P$-$\delta$. (b) the response rate analysis of $H_{sync}(s)$ by using its dominant $\mathrm{Re}\{\lambda\}$.

*2) Impact of Response Rate Condition on the Convergence*

According to the singular perturbation theory, $H_{sync}(s)$ can be regarded as in its steady-state only if its response is sufficiently fast than the characteristic timescale of the angle movement (typically four times smaller). In which, the response rate of $H_{sync}(s)$ can be evaluated by $\mathrm{Re}\{\lambda\}$, and the timescale of the synchronization (i.e., determines how fast the angle moves) can be generally emulated by the driven frequency $\Delta\omega$, for instance, a small $\Delta\omega$ indicates a slow angle movement and a relatively large timescale of the synchronization control.

Based on this, effects of the response rate on the convergent behavior are tested under the same synchronization timescale (i.e., $\Delta\omega$ = 0.06 p.u.). The results are shown in Fig. 3(b). Although all cases have stable poles and non-divergent PACs, their convergence quality varies. As $\mathrm{Re}\{\lambda\}$ of $H_{sync}(s)$ moves away from the imaginary axis, the convergence quality improves due to the faster response rate of $H_{sync}(s)$, which better satisfies the timescale decoupling condition. Otherwise, a non-convergent phenomenon may occur as the first row of Fig. 3(b). It is noted that for this case the stability condition of the $H_{sync}(s)$ is satisfied. This analysis highlights the importance of the response rate in ensuring the quasi-steady-state assumption of the PAC and achieving well-behaved convergence.

## IV. Conclusion

The VA-PAC derived from quasi-steady-state assumptions is a crucial tool for the current GFM converters' TSS analyses. However, the conditions under which it holds lack exposure. By developing a full-order VA-PAC model and performing an eigen-sweep-based analysis, this letter formally reveals a non-uniform convergent issue of the existing VA-PAC. The stability and response rate requirements for the satisfaction of the quasi-steady-state declaration of the VA-PAC are clarified. This work prompts an important issue regarding the need to re-examine the validity boundaries of existing PAC-based analyses, which may give rise to a series of analyses on the applicability of existing TSS analyses for GFM converters, and potentially lead to new findings and methods.